\documentclass[11pt,a4paper]{article}

\usepackage{jinstpub}

\usepackage[T1]{fontenc}
\usepackage[utf8]{inputenc}
\usepackage[english]{babel}
\usepackage{textcomp}

\usepackage{siunitx}
\usepackage{microtype}
\usepackage[capitalise,noabbrev]{cleveref}
\usepackage{placeins}

\usepackage{tikz}

\notoc

\begin{document}

% ---- Line numbers switch (draft only) ----
% \linenumbers
% \modulolinenumbers[5]

\title{Development and validation of a forward 0.7--4~MeV quasi-monoenergetic neutron capability at the CN Van de Graaff of LNL}

\author[a,b,1]{Jeffery~Wyss\note{Corresponding author.}}
\author[d]{Pierfrancesco~Mastinu}
\author[d]{Elizabeth~Musacchio~González}
\author[d]{Guido~Martín~Hernández}
\author[a,c]{Luca~Silvestrin}
\author[d]{Alberto~Monetti}
\author[e]{Hans~Th.\ J.~Steiger}
\author[e]{Manuel~B\"ohles}
\author[f]{B.~Lalremruata}
\author[g,h]{Saulo~Gabriel~Alberton}
\author[i]{David~Flechas}

\affiliation[a]{INFN -- Sezione di Padova, 35131 Padova, Italy}
\affiliation[b]{DICEM, University of Cassino and Southern Lazio, 03043 Cassino, Italy}
\affiliation[c]{University of Padova, 35131 Padova, Italy}
\affiliation[d]{INFN -- Laboratori Nazionali di Legnaro (LNL), 35020 Legnaro (PD), Italy}
\affiliation[e]{Technical University of Munich (TUM), Department of Physics, 85748 Garching, Germany}
\affiliation[f]{Department of Physics, Mizoram University, Aizawl 796004, India}
\affiliation[g]{Departamento de Fisica Nuclear, Instituto de Fisica, Universidade de S\~ao Paulo, 05508-090 S\~ao Paulo-SP, Brazil}
\affiliation[h]{``Gleb Wataghin'' Institute of Physics, State University of Campinas, 13083-859 Campinas-SP, Brazil}
\affiliation[i]{Departamento de Fisica, Universidad Nacional de Colombia, 111321 Bogota, Colombia}

\emailAdd{wyss@unicas.it}

\abstract{
The CN Van de Graaff accelerator of INFN--LNL provides forward-angle quasi-monoenergetic neutrons in the 0.7--4~MeV range via the
$^7$Li(p,n)$^7$Be reaction on thin metallic lithium targets. This work describes the development and experimental validation of this forward neutron capability, combining comparisons of commonly used transport tools with time-of-flight (ToF) measurements.

Neutron yields calculated with EPEN, FLUKA, MCNPX, and PINO are compared over the CN energy range in order to assess model-dependent variations relevant for fluence estimates. For zero incident-energy spread, a mutually consistent set of transport calculations agrees within $\pm5\%$ and is used as a practical reference for normalisation. The effect of incident-energy convolution on the predicted yields is examined.

Time-of-flight measurements performed using a sub-nanosecond secondary pulsing system verify the timing structure and forward-angle kinematics of the quasi-monoenergetic neutron component at the detector position, with neutron arrival times consistent with the expected forward kinematics within the experimental resolution.

Using measured proton currents and transport calculations based on this reference set, forward neutron fluences at the device position are estimated with an overall uncertainty of approximately $9\%$, including contributions from current integration, target thickness, and geometry.

A short device irradiation, carried out in parallel with the ToF campaign, demonstrates measurable response under CN beam conditions and confirms the practical usability of the beam for low-MeV neutron studies. 
Together, these results establish the current operational performance of the CN $0^\circ$ forward quasi-monoenergetic neutron capability in the 0.7--4~MeV range and identify the steps required toward routine calibrated operation.
}

\keywords{Quasi-monoenergetic neutrons, Time-of-flight measurements, $^{7}$Li(p,n)$^{7}$Be reaction, Radiation effects in microelectronics, single-event effects, single-event upset, soft errors}

% PACS is not part of the standard JINST front matter; if you must keep it, leave it here:
% \PACS{28.20.-v, 28.20.Cz, 25.40.-h, 29.25.Dz, 29.30.Hs, 85.40.-e}

\maketitle
\raggedbottom
\flushbottom
\clearpage

%---------------------- INTRODUCTION ----------------------
\section{Introduction}

Quasi-monoenergetic (QMN) neutron beams produced via the
$^7$Li(p,n)$^7$Be reaction are widely used for detector development, nuclear-data studies, and validation of transport calculations. In the neutron energy range below about 5~MeV, noticeable differences persist among commonly used reaction and transport tools, reflecting variations in the treatment of reaction channels and incident-energy convolution. Experimental verification of forward-angle neutron production in this regime is therefore important for establishing reliable working reference fields.

The CN Van de Graaff accelerator of INFN--LNL provides forward
quasi-monoenergetic neutrons in the 0.7--4~MeV range using thin metallic lithium targets. With the addition of a secondary pulsing system, the beam can be delivered in sub-nanosecond bunches suitable for time-of-flight (ToF) measurements. This configuration enables direct checks of the forward-angle neutron timing structure and provides access to a controlled low-MeV neutron field suitable for detector and radiation-effects studies.

The present work describes the development and validation of the CN
$0^\circ$ forward QMN capability under current operating conditions. Neutron yields from several commonly used transport tools are compared to assess model-dependent variations and define a consistent reference set for fluence estimates. Time-of-flight measurements are then used to verify the timing structure and forward-angle kinematics of the quasi-monoenergetic neutron component at the experimental station.

From these elements, the achievable forward neutron fluence and its associated uncertainties are determined. A brief irradiation of a memory device was performed as a simple single-event upset (SEU) test to check the practical usability of the beam. The primary focus of this paper, however, is the definition of the neutron-field performance itself rather than a detailed study of device response.

%================== SECTION 2: ACCELERATOR ==================
\section{The accelerator}

The CN Van de Graaff accelerator at LNL \cite{CN} provides
low-energy proton and deuteron beams with terminal voltages up to
6\,MV, and can reach approximately 6.5\,MV under suitable conditioning. Continuous proton currents up to about 6\,\textmu A are technically achievable, while in pulsed mode up to 700\,nA can be delivered at a repetition rate of 3\,MHz.

In practice, the usable proton current is constrained by
radiation-protection requirements associated with neutron and gamma
production in the switching-magnet and target areas. The authorised
continuous-beam limit is presently about 4\,\textmu A, although lower limits apply depending on the reaction and target configuration. For proton irradiation of thin lithium targets used for neutron production, the effective current is significantly reduced by radiological constraints at the target station.

These current limits set the available forward neutron fluence and
therefore determine the achievable neutron rates at the experimental position.

%================== SECTION 3: MODELLING ==================
\section{Neutron-yield modelling and comparison}
\label{sec:modelling}

This section compares neutron yields for the $^{7}$Li(p,n)$^{7}$Be reaction at CN energies and examines the agreement among commonly used transport tools for forward-angle yield predictions. From this comparison, a model-dependent spread is defined that enters the uncertainty on the fluence estimate. The modelling is presented first in order to define a reference forward-yield and its associated uncertainty, which is subsequently used for the interpretation of the experimental results.

The $^7$Li(p,n)$^7$Be reaction is widely used for accelerator-based neutron production owing to its comparatively low threshold energy ($E_{\text{thres}} = 1.9$~MeV) and favourable yield. At CN beam energies, the forward-emitted neutrons span approximately 0.7--4~MeV. For proton energies below $\sim$3.7~MeV, the forward spectrum is dominated by the two-body $^7$Li(p,n$_0$)$^7$Be and $^7$Li(p,n$_1$)$^7$Be channels, corresponding respectively to neutron production with the residual $^7$Be nucleus in its ground state and first excited state. These channels produce a forward quasi-monoenergetic structure whose width is determined mainly by proton energy loss in the lithium target. Above $E_p \sim 3.7$~MeV, the opening of three-body breakup channels introduces an additional broader low-energy continuum component, leading to progressive broadening of the forward neutron spectrum. The excitation function exhibits a narrow resonance at $E_p \approx 2.2$--$2.3$~MeV and a broader maximum near 5~MeV~\cite{Bertini,experimental_7Be_cross}.

\begin{figure}[!htbp]
    \centering
    \includegraphics[width=\linewidth]{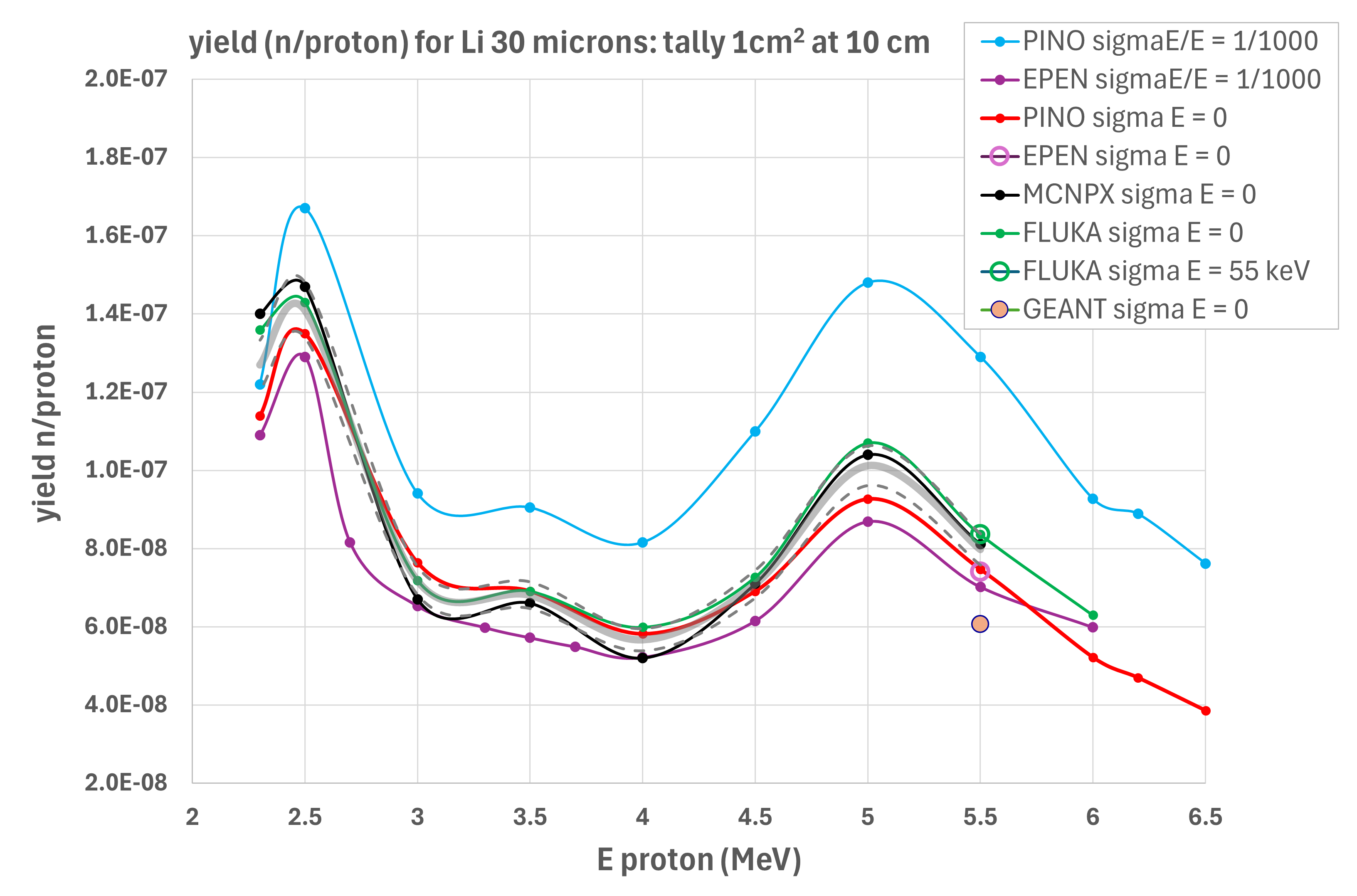}
\caption{
Forward neutron yield per incident proton predicted by different simulation tools for protons on a 30~\textmu m Li target (1\,cm$^2$ tally area at 10~cm distance). Results from EPEN, FLUKA, MCNPX, GEANT4 (all with $\sigma_E = 0$ or very small), and PINO ($\sigma_E = 0$ and $\sigma_E = 5.5$~keV) are shown. The grey dashed curves indicate the upper and lower bounds of the $\pm5\%$ interval constructed around the mutually consistent zero-spread calculations from EPEN, FLUKA, and MCNPX.
}
    \label{fig:comparison_yields}
\end{figure}

Despite the long-standing use of this reaction, differences persist among transport tools in the prediction of absolute forward yields. Calculations were performed using PINO~\cite{PINO}, GEANT4~\cite{geant}, FLUKA~\cite{fluka}, MCNPX~\cite{MCNPX}, and the deterministic code EPEN~\cite{EPEN}. All tools reproduce the general spectral structure, but small differences in integral yields are observed (Fig.~\ref{fig:comparison_yields}). This comparison shows that, in the zero-spread limit, a consistent set of tools defines a stable reference for forward-yield normalisation.

At zero incident-energy spread, EPEN, FLUKA, and MCNPX define a central reference forward-yield curve, with mutually consistent predictions over the CN energy range. The grey dashed curves in Fig.~\ref{fig:comparison_yields} indicate the upper and lower bounds of the $\pm5\%$ interval constructed around this reference curve. This interval is used as the model-dependent uncertainty for flux normalisation. PINO results obtained with $\sigma_E = 0$ lie within these bounds.

When a finite incident-energy spread is introduced in PINO ($\sigma_E/E \sim 10^{-3}$), the predicted yield increases relative to the zero-spread result. An equivalent sensitivity is not observed in EPEN or FLUKA under the same imposed spread. Because the $^7$Li(p,n)$^7$Be cross section varies approximately linearly over a $\pm1\%$ interval around $E_p = 5.5$~MeV, a symmetric incident-energy spread is not expected to modify the mean forward yield significantly. The enhanced yield predicted by PINO for $\sigma_E>0$ therefore suggests a stronger sensitivity to the treatment of incident-energy convolution than is observed in the corresponding EPEN or FLUKA calculations, rather than a difference arising from the underlying nuclear data themselves.

\begin{figure}[!htbp]
    \centering
    \includegraphics[width=\linewidth]{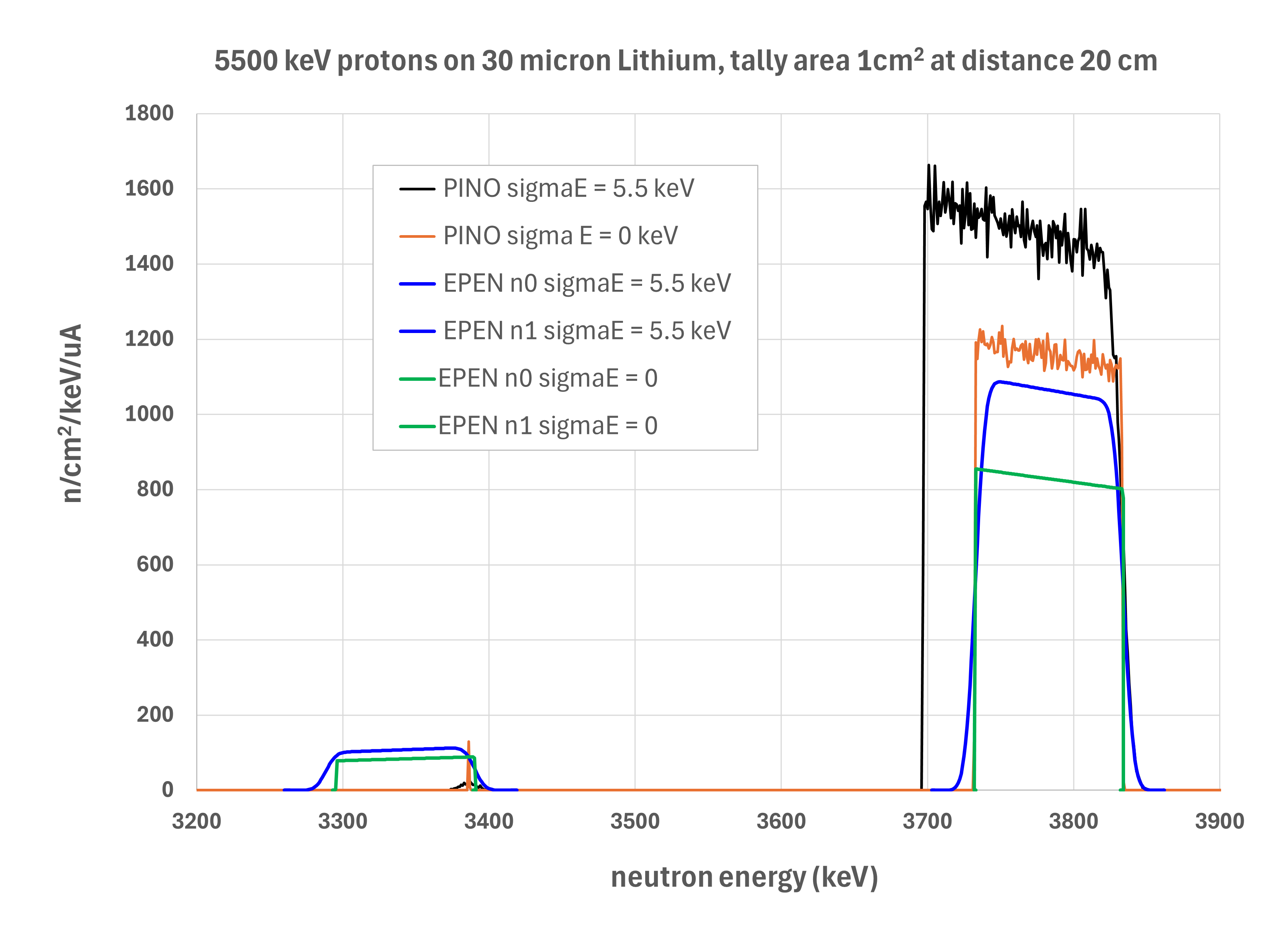}
    \caption{
Differential neutron spectra at 0$^\circ$ for 5.5~MeV protons on a 30~\textmu m Li target (tally: 1\,cm$^2$ at 20~cm). Results are shown for the (p,n$_0$) and (p,n$_1$) channels from EPEN and PINO, with and without an assumed incident-energy spread ($\sigma_E=5.5$~keV).}
    \label{fig:PINO_EPEN_5500_keV}
\end{figure}

Figure~\ref{fig:PINO_EPEN_5500_keV} compares representative forward-angle spectra. While integral yields may agree among codes in the zero-spread limit, differences in the relative contributions of the (p,n$_0$) and (p,n$_1$) channels remain, and none of the tools fully model the three-body breakup channel above 3.7~MeV. This illustrates that agreement in integral yield does not imply consistency in the underlying spectral composition. Figure~\ref{fig:PINO_EPEN_5500_keV} is therefore intended primarily as an illustration of the relative spectral structure and convolution sensitivity of the different calculations, rather than as a benchmark of absolute spectral accuracy.

\begin{figure}[!htbp]
    \centering
    \includegraphics[width=\linewidth]{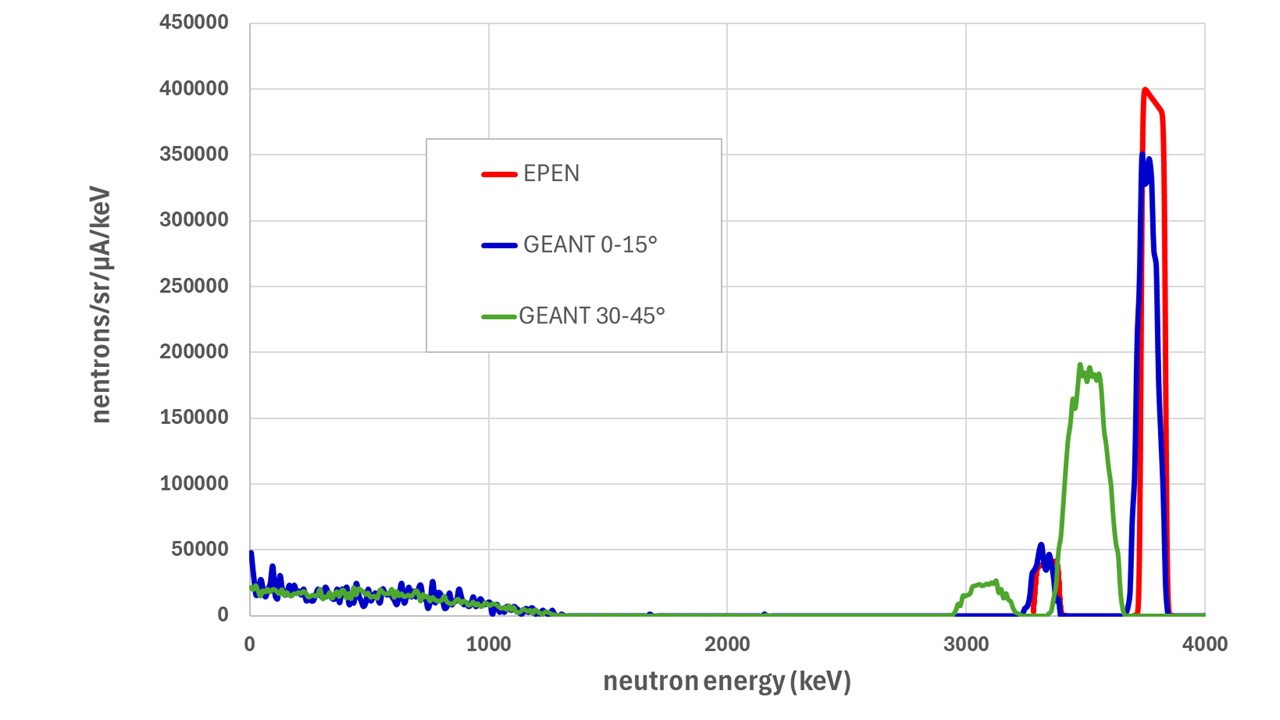}
    \caption{
Comparison of GEANT4 and EPEN spectra for 5.5~MeV protons on a 30~\textmu m Li target with a 300~\textmu m Cu backing (tally: 1\,cm$^2$ at 20~cm).}
    \label{fig:new_EPEN_GEANT}
\end{figure}

GEANT4 and EPEN spectra including a 300~\textmu m copper backing are compared in Fig.~\ref{fig:new_EPEN_GEANT}. Neutrons below 1.4~MeV originate primarily from the copper stopper and disappear when the stopper is removed in simulation, illustrating the sensitivity of the low-energy tail to beam-line materials. This demonstrates that the low-energy component of the spectrum is strongly influenced by beam-line materials and must be accounted for when interpreting the forward neutron field.

These comparisons show that forward neutron yields in the CN energy range are reproducible within approximately $\pm5\%$ across the considered transport tools when zero incident-energy spread is assumed. This spread is used as a practical model-dependent uncertainty for subsequent fluence estimates. While the agreement among codes does not in itself guarantee absolute accuracy, since common nuclear data libraries or similar modelling assumptions may be shared, it nevertheless provides a practical and internally consistent reference for forward-yield normalisation in the absence of a dedicated experimental benchmark.

%========= SECTION 4: EXPERIMENTAL SYSTEMS ==================
\section{Experimental systems}
\label{sec:experimental}

This section describes the hardware elements required to operate the CN beam line as a neutron source capable of supporting time-of-flight (ToF) measurements for consistency checks and soft-error testing.

\subsection{Lithium target system}

\begin{figure}[t]
    \centering
    \includegraphics[width=\linewidth]{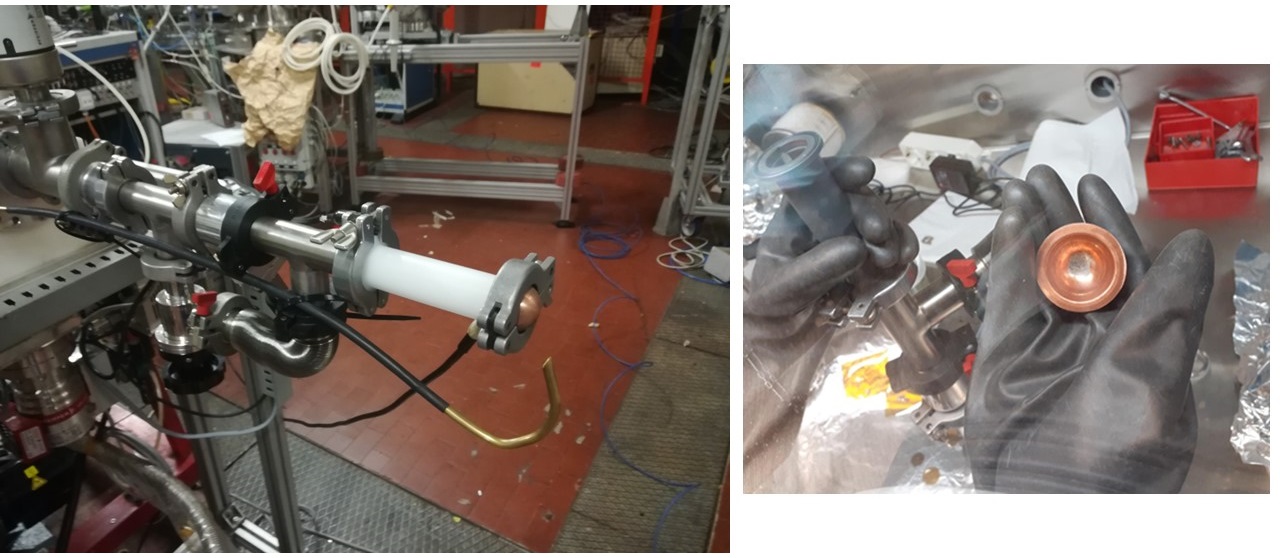}
    \caption{
Air-cooled 300~\textmu m copper beam stopper used at the end of the 0$^\circ$ beam line (left) and close-up view of the pressed metallic lithium layer mounted on its front face (right). Lithium foils of 10--40~\textmu m thickness are fabricated inside an argon-filled glovebox to prevent oxidation.
}
    \label{fig:CN_target}
\end{figure}

The neutron-production target consists of a metallic lithium layer pressed into a 300~\textmu m-thick copper (Cu) backing that also serves as the beam stopper (Fig.~\ref{fig:CN_target}). In its present configuration, the copper-backed target can dissipate up to 10~W, corresponding to maximum continuous beam currents of 5~\textmu A at 2~MeV and 1.8~\textmu A at 5.5~MeV, whereas a tantalum beam stopper would tolerate only $\sim$4~W.

\begin{figure}[t]
    \centering
    \begin{tikzpicture}
        \node[anchor=south west, inner sep=0] (img) at (0,0)
            {\includegraphics[width=\linewidth]{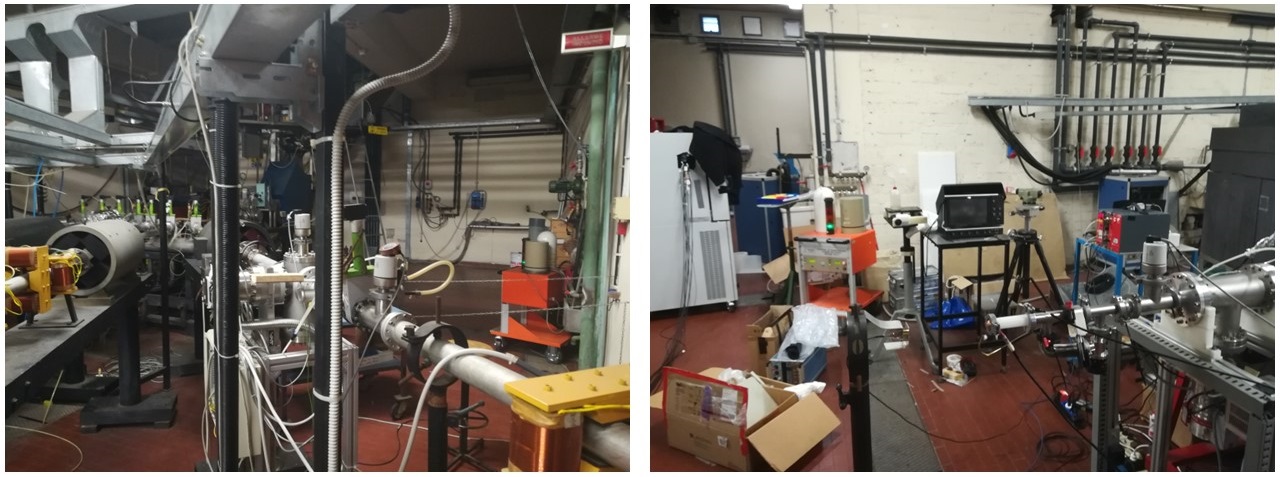}};
        \begin{scope}[x={(img.south east)},y={(img.north west)}]
            % Adjust coordinates after one compile
            \draw[red, thick] (0.42,0.42) ellipse (0.05 and 0.11);
            \node[red, fill=white, inner sep=1pt, anchor=west] at (0.35,0.57) {\scriptsize Limiting monitor};

            \draw[red, thick] (0.665,0.50) ellipse (0.05 and 0.12);
            \node[red, fill=white, inner sep=1pt, anchor=west] at (0.61,0.66) {\scriptsize Hall monitor};
        \end{scope}
    \end{tikzpicture}
    \caption{
Neutron monitors used to enforce the 50~\textmu Sv/h radiation-protection limit.
The limiting monitor near the switching magnet (left) defines the operational beam-current constraint, while the hall monitor near the 0$^\circ$ target assembly (right) verifies that radiation levels at the experimental position remain within acceptable limits.
}
    \label{fig:neutron_monitors}
\end{figure}

\begin{figure}[!htbp]
    \centering
    \includegraphics[width=0.8\linewidth]{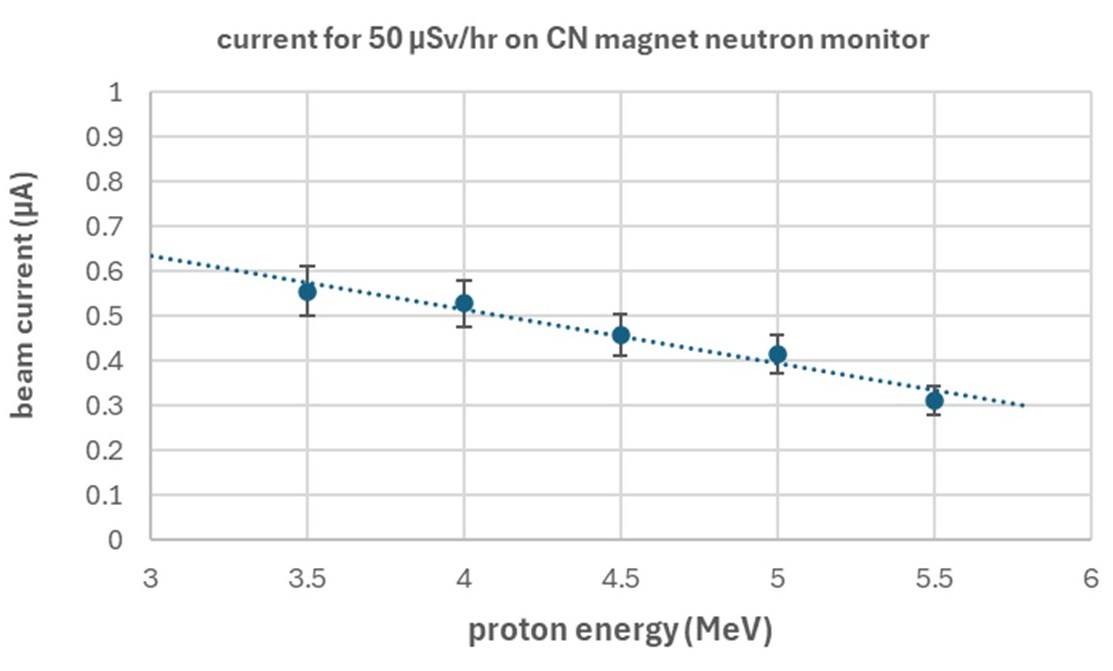}
    \caption{
Approximate proton-beam current as a function of energy corresponding to the 50~\textmu Sv/h threshold measured by the neutron monitor next to the bending magnet.}
    \label{fig:limit_current}
\end{figure}

In the present configuration, the operational limit is set not by target heating but by radiation-protection thresholds. The most restrictive constraint is the 50~\textmu Sv/h limit imposed by a neutron monitor located near the switching magnet and operator shielding (Fig.~\ref{fig:neutron_monitors}, left). For 5.5~MeV protons, this limits the continuous current to $\sim$0.3~\textmu A. At the same current, the neutron monitor located closer to the target (Fig.~\ref{fig:neutron_monitors}, right) measures $\sim$450~\textmu Sv/h, well below the 3~mSv/h limit for the experimental hall. For lower proton energies, the threshold is reached at proportionally higher currents (Fig.~\ref{fig:limit_current}). The monitor near the switching magnet therefore defines the operational current limit, while the hall monitor provides a check that radiation levels at the experimental station remain within acceptable bounds.

A modest additional local shielding package (approximately 5--10\,cm of borated polyethylene plus lead) around the switching magnet and target zone is under evaluation. This upgrade is expected to reduce the dose rate at the limiting monitor by a factor of 4--6 and allow operation at continuous proton currents of order 1\,\textmu A across the 0.7--4\,MeV range (and up to $\sim$2--3\,\textmu A below 2\,MeV) within the existing radiological limits.

Metallic lithium has low mechanical strength and requires a rigid support, provided here by the copper backing. Because of its chemical reactivity, lithium is handled under inert atmosphere; fabrication is performed inside an argon-filled glovebox from lithium stored under paraffin. The lithium is cleaned and rolled to the desired thickness, typically 20~\textmu m (with thinner foils down to sub-10~\textmu m), and pressed onto the copper backing using paired moulds.

\begin{figure}[!htbp]
    \centering
    \includegraphics[width=\linewidth]{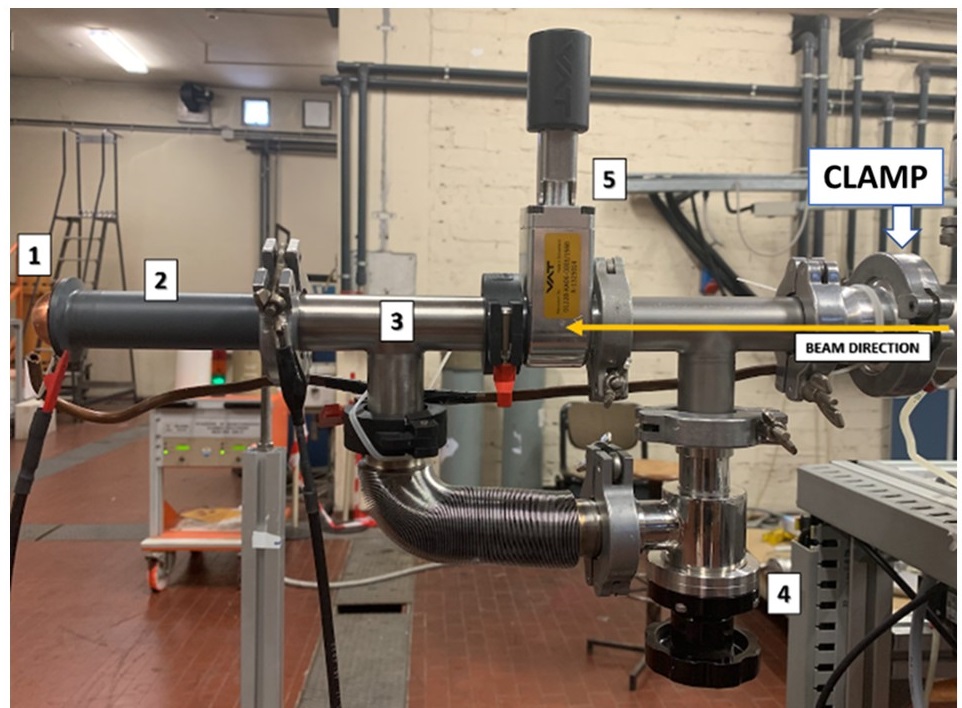}
    \caption{
The lithium target assembly at the end of the beam line:
(1) removable Cu backing; (2) isolated PVC tube; (3) Ta collimator;
(4) micrometric valve; (5) second gate valve.
}
    \label{fig:target_system}
\end{figure}

\paragraph{Thermal analysis}

An axisymmetric finite-element (FE) thermal analysis was performed assuming a uniform 10~W heat load over a 3~mm beam spot in perfect thermal contact with the Cu capsule. Heat dissipation through the clamp and gasket and radiative losses were neglected.

\begin{figure}[t]
    \centering
    \includegraphics[width=\linewidth]{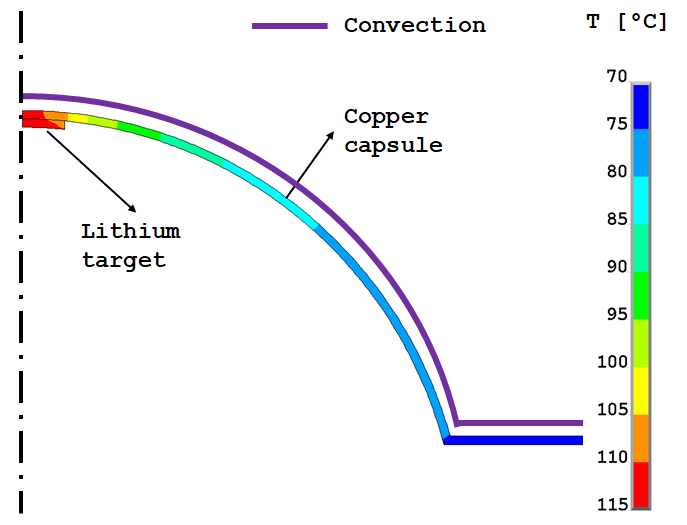}
    \caption{
Boundary conditions and calculated temperature distribution for 10~W deposited on the lithium target encapsulated by the Cu beam stopper.
}
    \label{fig:target_temp}
\end{figure}

A convection coefficient of 100~W\,m$^{-2}$\,K$^{-1}$ was applied to the external Cu surface. Under these assumptions, the lithium temperature rises by approximately 6$^\circ$C per watt from convection and a further 4$^\circ$C per watt from radial conduction (Fig.~\ref{fig:target_temp}). The target can safely dissipate $\sim$10~W, corresponding to continuous proton currents of 5~\textmu A at 2~MeV or 1.8~\textmu A at 5.5~MeV; increasing the Cu thickness to 600~\textmu m would raise the allowable power by $\sim$25\%.

\subsection{Secondary pulsing system for the 0$^\circ$ beam line}

The secondary pulsing system was developed to provide sub-nanosecond timing for the ToF measurements described in the next section.

The CN Van de Graaff accelerator can operate in continuous-wave or pulsed mode. In the standard pulsed configuration, the beam frequency is 3~MHz (333~ns period). A buncher compresses the proton pulses in time at the cost of increased energy spread, yielding pulse widths below 2~ns (FWHM).

The secondary pulsing system suppresses selected pulses from the 3~MHz beam, producing repetition frequencies between 1.5~MHz and 500~kHz. The resulting pulses have sub-nanosecond width ($\simeq$1~ns FWHM, down to 0.9~ns) and remain phase-locked to the accelerator frequency.

\begin{figure}[t]
    \centering
    \includegraphics[width=\linewidth]{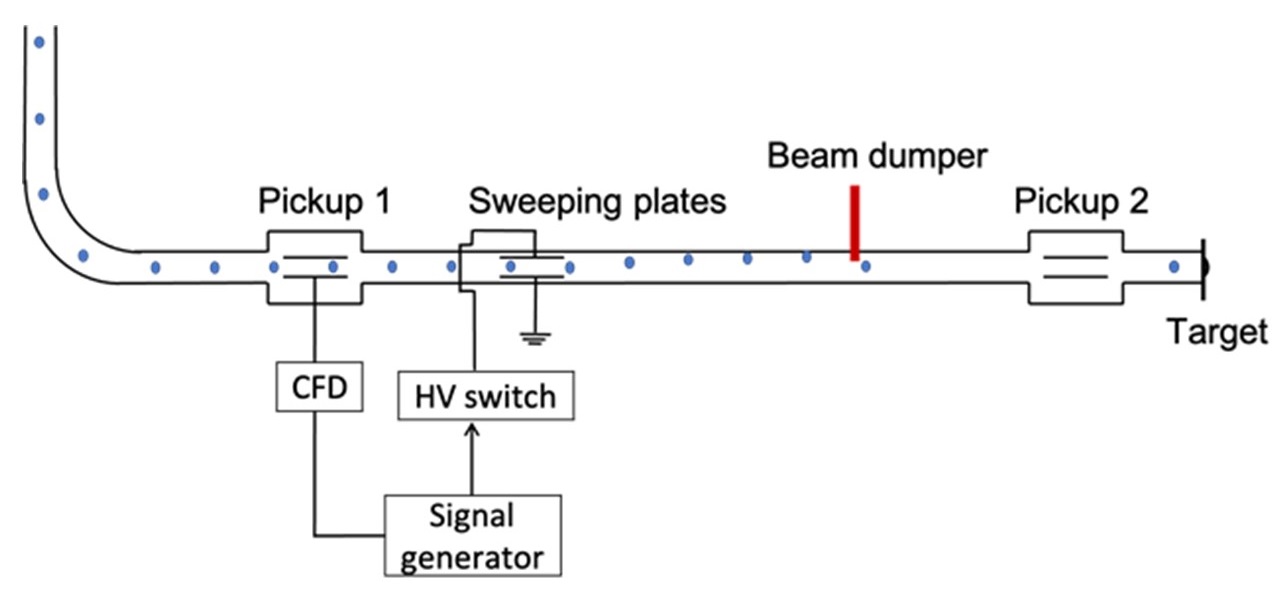}
    \caption{Schematic layout of the secondary pulsed-beam system.}
    \label{fig:kicker}
\end{figure}

A schematic is shown in Fig.~\ref{fig:kicker}. A pair of aluminium electrodes generates the transverse electric field used to deflect the beam. One electrode is grounded, while the other is switched by a fast solid-state module (Behlke GmbH~\cite{behlke}). The switching signal is derived from a constant-fraction discriminator triggered by a capacitive pickup.

\begin{figure}[t]
    \centering
    \includegraphics[width=0.75\linewidth]{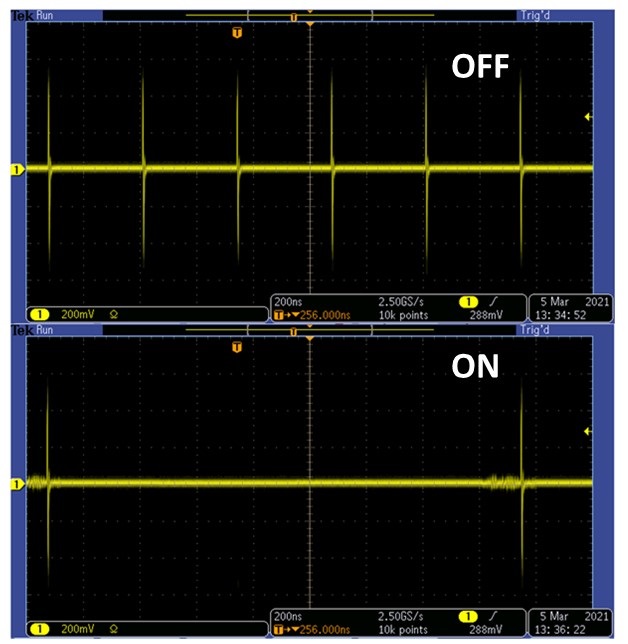}
    \caption{
Oscilloscope traces from Pickup~2.
Top: main 3~MHz proton beam.
Bottom: 600~kHz proton beam after activation of the secondary pulsing system.
}
    \label{fig:kicker_scope}
\end{figure}

The system, developed at LNL and later adopted by NEC~\cite{NEC}, required minimisation of load capacitance to maintain fast switching. The achieved 0.9~ns FWHM pulse width demonstrates that the system meets the ToF requirements at the CN beam line.

We now turn to experimental verification of the timing structure and forward-angle kinematics of the neutron field using the ToF technique.
\FloatBarrier

%================== SECTION 5: TOF VALIDATION ==================
\section{Time-of-flight verification of the QMN component}
\label{sec:tof_validation}

Accurate knowledge of the QMN neutron spectra produced on the CN~0$^\circ$ beam line is important for interpreting the simulations discussed in Section~\ref{sec:modelling} and for assessing the facility’s performance. The secondary pulsing system provides the timing structure required for direct ToF measurements, enabling verification of the prompt timing features and forward-angle kinematics at the detector position.

\subsection{Liquid-scintillator ToF measurements}

The 0$^\circ$ beam line, equipped with thin lithium targets and the secondary pulsing system, is used to study neutron and $\gamma$ responses in liquid scintillators (LS) developed for large neutrino-physics and rare-event experiments~\cite{Juno,Hans,WbLS}. Pulse-shape analysis enables neutron--gamma discrimination, allowing clean separation of recoil-proton and Compton events under identical beam conditions.

\begin{figure}
    \centering
    \includegraphics[width=\textwidth]{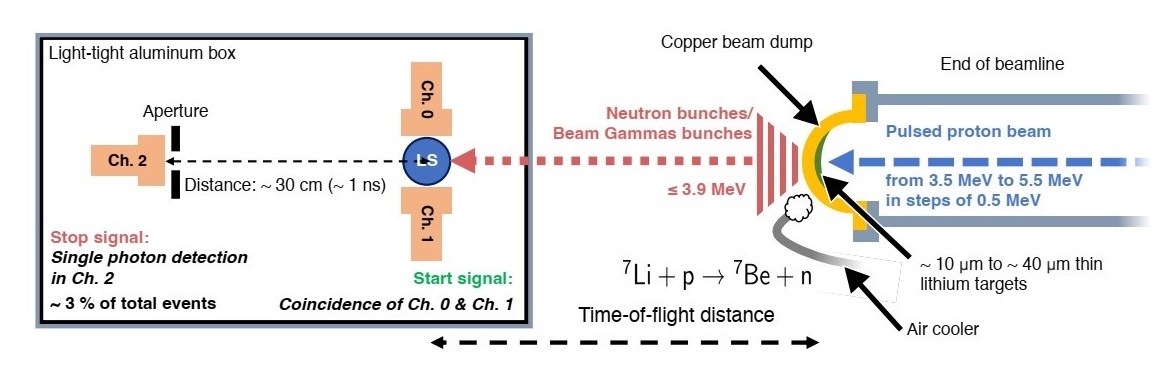}
    \caption{
Schematic of the liquid-scintillator setup used for scintillation-timing and neutron ToF measurements.}
    \label{fig:TOF_setup}
\end{figure}

A schematic of the experimental arrangement is shown in Fig.~\ref{fig:TOF_setup}. The LS vessel is viewed by two PMTs providing a coincidence start signal with sub-nanosecond precision. A third PMT, positioned 60~cm behind a small aperture, operates in single-photon mode in a TCSPC configuration for the optical stop signal. The neutron ToF start is derived from the proton-beam pickup located upstream (Fig.~\ref{fig:kicker}), enabling simultaneous neutron--gamma discrimination and neutron ToF measurements.

\begin{figure}[htbp]
 \centering
 \includegraphics[width=\linewidth]{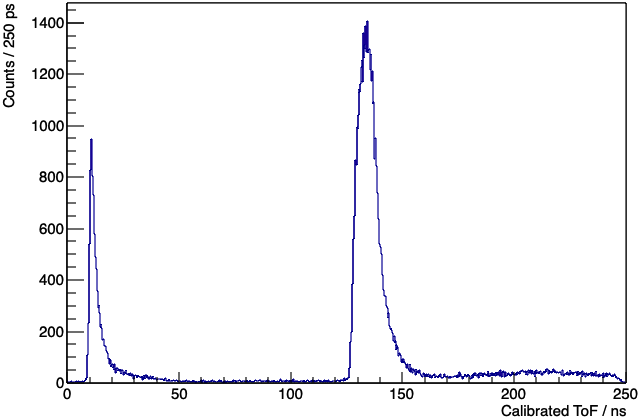}
\caption{
Representative ToF distribution for $E_p = 5.5$~MeV protons on a 20~\textmu m Li target, with the detector at 3.46~m and 0$^\circ$. The prompt $\gamma$ flash defines $t=0$, while the delayed peak at $\sim$130~ns corresponds to the forward neutron component.
}
 \label{fig:tof}
\end{figure}

Pulsed proton beams at $\sim$600~kHz and energies between 3.5 and 5.5~MeV ($\sigma_E\sim$3~keV) were directed onto 10--40~$\mu$m Li targets, producing QMN neutrons with mean energies of 1.7--3.8~MeV. Figure~\ref{fig:tof} shows a calibrated ToF distribution recorded for 5.5~MeV protons on a 20~$\mu$m Li target. The prompt $\gamma$ flash at $t=0$ is followed by the neutron peak at $\sim$130~ns, corresponding to the 3.46~m flight path.

The asymmetric shape of the neutron peak reflects the forward-angle $^7$Li(p,n)$^7$Be kinematics. 
The quasi-flat neutron energy distribution produced by proton energy loss in the lithium target maps through the non-linear time-of-flight relation into a correspondingly skewed time distribution. The peak width is dominated by neutron-production kinematics and target-related energy spread rather than by the $\sim1$~ns beam pulse width or detector timing.

For neutrons with $E_n=1$~MeV, the arrival time is $\sim$230~ns; a 245~ns acquisition gate excludes slower, sub-MeV components and late room-return. The $n_0$ and $n_1$ two-body channels are not resolved at this flight path and contribute to a single composite peak.

The slower neutrons arriving after the main peak arise from several contributions. At this energy, the $^7$Li$(p,n^3)^4$He breakup channel produces neutrons with $E_n \leq 1.4$~MeV peaking near 0.6~MeV~\cite{1972paper}. Additional sub-MeV neutrons originate from the copper beam stopper and from room-return.

Forward-angle neutron production from copper can be eliminated at the highest CN proton energies by inserting a thin gold layer between the lithium and copper. For $E_p \leq 4$~MeV the Cu backing cannot produce (p,n) neutrons, and below $E_p \simeq 3.7$~MeV the three-body breakup channel is closed.

The ToF distribution in Fig.~\ref{fig:tof} therefore verifies the timing structure and forward-angle kinematics of the QMN component. At this stage, the ToF data are used as a timing and kinematic validation; quantitative spectral reconstruction is deferred to future work.

Although the present ToF analysis is not intended as a quantitative spectral unfolding, it is still useful to check whether the observed event rates are broadly consistent with the forward neutron yields adopted for the fluence estimates. In this context, a rate-level consistency check was performed at $E_p = 5.5$~MeV with the detector at 2.85~m using a $75\times75$~mm$^2$ acceptance and a 20~$\mu$m lithium target.

For this geometry, PINO predicts a forward neutron rate of $2.14\times10^{6}$~s$^{-1}$ per 100~$\mu$A for $\sigma_E=0$. In this configuration, PINO agrees within $\pm5\%$ with EPEN, FLUKA, and MCNPX, as discussed in Section~\ref{sec:modelling}.

After scaling to the respective beam currents (60~nA and 37.5~nA), the two runs yield consistent fast-neutron fractions of $\sim$0.43 under the applied ToF conditions. This agreement indicates that at 5.5~MeV the rate-level response is governed primarily by geometry and ToF selection.

A full spectral unfolding and quantitative comparison with transport simulations are beyond the scope of the present work. In particular, a direct comparison between the measured time-of-flight distributions and transport-model predictions would require inclusion of detector response and efficiency effects, together with proper treatment of the non-linear relation between neutron energy and time-of-flight. A quantitative comparison would also require transforming the simulated neutron spectra into the time domain. These effects are not included in the present analysis. The present ToF measurements are therefore limited to validation of the timing structure and forward-angle kinematics of the neutron field.

These results establish the timing structure and forward-angle kinematics of the neutron field. We now turn to a practical demonstration of beam usability through a feasibility-level SEU measurement.

\FloatBarrier

%==== SECTION 6: propects and feasibility =========
\section{SEU feasibility study at CN}
\label{sec:SEU_merged}

The forward quasi-monoenergetic neutron beam at CN provides access to the 0.7--4~MeV energy range, relevant for exploratory single-event upset (SEU) studies in modern electronic devices, a region where experimental data remain comparatively limited. To assess the operational viability of detecting SEU activity, we performed a short parasitic measurement under the current beam conditions.

For context, Table~\ref{tab:facilities} compares the CN beam line with a selection of accelerator-based low-energy neutron sources relevant to detector development, nuclear-data studies, and exploratory SEE work. The values are indicative, based on publicly available documentation, and are included only to place the CN 0.7--4~MeV capability in the broader low-energy landscape.

\begin{table}[htbp]
\centering
\footnotesize
\caption{
Indicative comparison of selected accelerator-based low-energy neutron sources relevant to detector development, nuclear-data work, and exploratory SEE studies. Energy ranges and forward-angle fluxes are approximate order-of-magnitude values taken from representative facility publications and public documentation~\cite{CN,UCL,PTB,TUD_ELBE}. This table is intended solely to contextualise the CN 0.7--4~MeV capability.
}
\label{tab:facilities}
\begin{tabular}{lccc}
\hline
Facility & Energy range & Flux at 0$^\circ$ & ToF mode \\
         & (MeV)        & (n\,cm$^{-2}$\,s$^{-1}$\,\textmu A$^{-1}$) & \\
\hline
LNL CN             & 0.7--4 (QMN)        & $\sim(2$--$8)\times10^{4}$ & Yes (ns pulsed) \\
UCL (Louvain)      & 1--6 (QMN)          & $\sim10^{5}$               & Limited (\textmu s pulsed) \\
PTB (Braunschweig) & 0.2--19 (QMN+broad) & $10^{3}$--$10^{5}$         & No (CW) \\
TUD-ELBE (Dresden) & 0.01--10 (white)    & $10^{4}$--$10^{6}$         & Yes (bunched) \\
\hline
\end{tabular}
\end{table}

A primary operational constraint is the available proton-beam current. Achieving useful SEU statistics requires at least 1~\textmu A of continuous CN current. This level should become attainable with the shielding improvements discussed in Section~\ref{sec:experimental}, enabling measurements of device cross sections in the $E_{\mathrm{n}}\sim0.7$--4~MeV region.  

Typical CN allocations for dedicated experiments are $\sim$10~days, of which $\sim$7~days are effectively available for data taking after setup and energy tuning. At present currents, the beam is sufficient for feasibility studies, whereas higher sustained currents are required for statistically meaningful cross-section measurements.

\begin{figure}[htbp]
    \centering
    \includegraphics[width=0.60\linewidth]{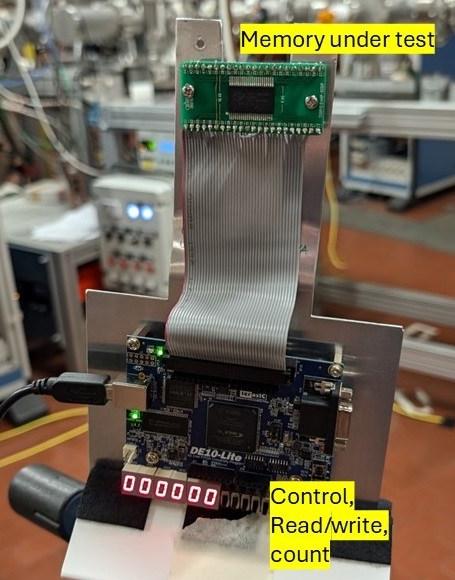}
    \caption{
FPGA-based SEU monitor used in the feasibility test. The control circuitry was not shielded against gamma radiation during this parasitic measurement.}
    \label{fig:SEU_monitor}
\end{figure}

Within this context, a short parasitic measurement was performed to assess the detectability of SEU activity under the available beam conditions.
The feasibility test was carried out using an FPGA-based memory-control system (Fig.~\ref{fig:SEU_monitor}), the detailed architecture of which is subject to a non-disclosure agreement. A TTL-level pulse is asserted on a dedicated FPGA pin whenever an error is detected by the on-board correction logic, and these pulses were counted electronically. The resulting counts were used to derive indicative device-level cross sections. 

The SEU monitor was operated continuously throughout the experimental campaign; no events were recorded during beam-off periods, indicating negligible background. No dedicated measurements were performed below the neutron-production threshold of the $^{7}$Li(p,n)$^{7}$Be reaction; however, no SEU activity was observed in the absence of beam, including during extended periods with the detector in place. This indicates that environmental background contributions (e.g. atmospheric neutrons) are negligible under the present conditions.

\begin{figure}[!h]
    \centering
    \includegraphics[width=0.7\linewidth]{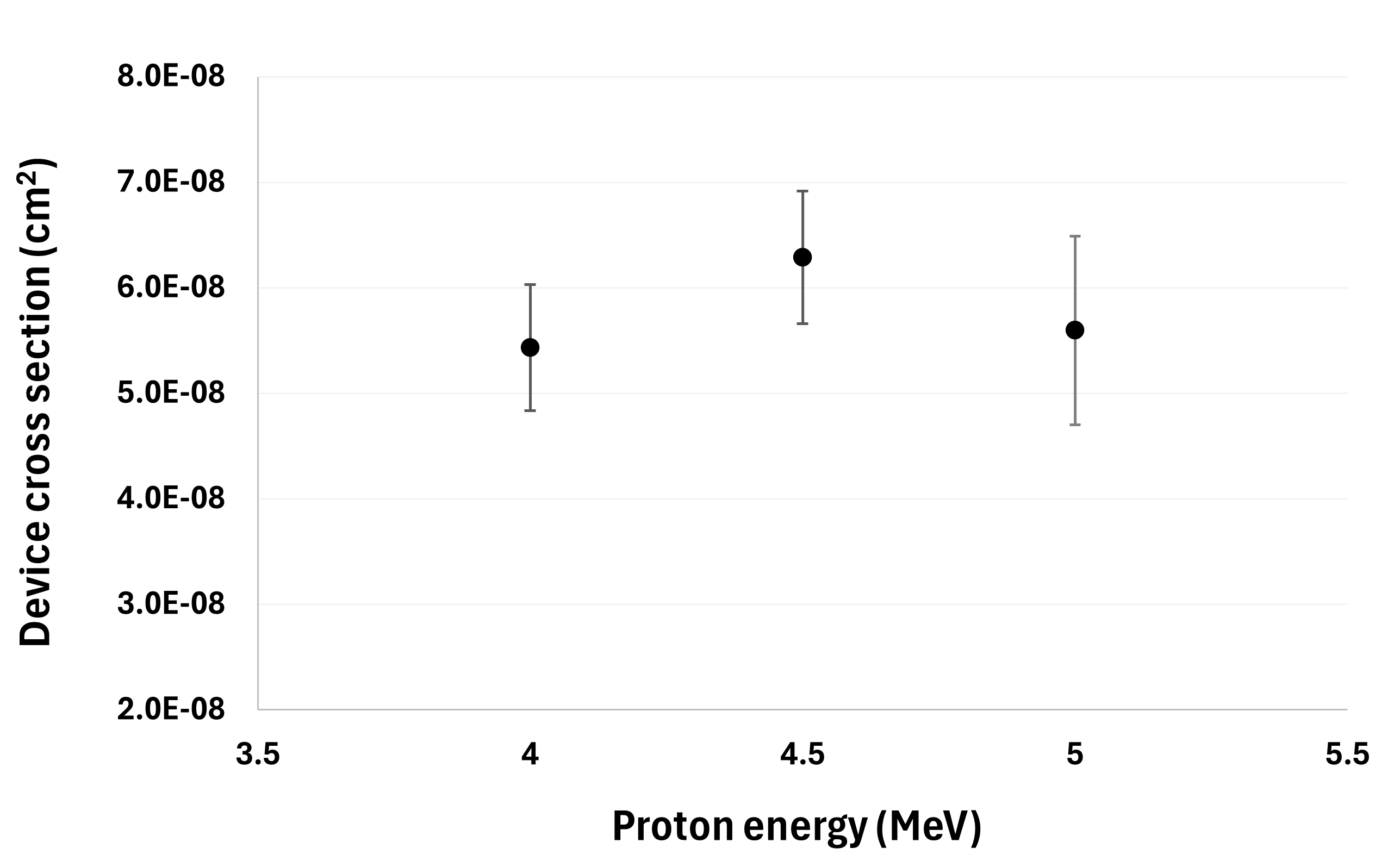}
\caption{
Device-level SEU cross sections for the proprietary memory device, derived from PINO-calculated neutron fluence at three CN proton energies. The cross sections are shown as a function of proton energy. Error bars include statistical and fluence-related systematic uncertainties. Within these uncertainties, the three measurements are mutually consistent.
}
\label{fig:CN_data}
\end{figure}

The test was performed parasitically during a liquid-scintillator ToF experiment at three proton energies (4.0, 4.5, and 5.0~MeV). Due to limited measurement time ($<5$~h), the geometry could not be stabilised; the DUT was repeatedly removed and repositioned with $\sim$1~mm reproducibility. The control electronics were unshielded. The memory was positioned 50~mm downstream of the target for the 4.0 and 5.0~MeV runs, and 40~mm for the 4.5~MeV run. High-statistics runs at 5.0 and 5.5~MeV were attempted, but total-dose artefacts appeared in the counting circuit, resulting in double counts; these runs were excluded from the quantitative analysis in Table~\ref{tab:CN_data} and Fig.~\ref{fig:CN_data}.

Device cross sections were calculated as
\begin{equation}
\label{eq:sigma_device}
\sigma_{\mathrm{device}} = \frac{N_{\mathrm{SEU}}}{N_{\mathrm{n}}},
\end{equation}
where $N_{\mathrm{SEU}}$ is the number of observed upsets and $N_{\mathrm{n}}$ is the PINO-calculated number of neutrons crossing the $1.5~\mathrm{cm}^{2}$ sensitive area of the DUT for the corresponding geometry and target configuration.

The individual run parameters and corresponding neutron numbers are summarised in Table~\ref{tab:CN_data}.

\begin{table}[htbp]
\centering
\footnotesize
\begin{tabular}{|c|c|c|c|c|c|c|c|}
\hline
$E_p$ & $I_p$ & Time & $N_p$ & Distance & Li thickness & $N_n$ (PINO, $\sigma_E=0$) & $N_{\text{SEU}}$ \\ 
(MeV) & (nA) & (min) & (protons) & (mm) & (\textmu m) & (through DUT area) & \\ 
\hline
4.0 & 430 & 36.5 & $5.88\times10^{15}$ & 50 & 40 & $2.76\times10^{9}$ & 102 \\
4.0 & 420 & 38.5 & $6.07\times10^{15}$ & 50 & 40 & $2.85\times10^{9}$ & 101 \\
4.5 & 340 & 117.6 & $1.50\times10^{16}$ & 50 & 20 & $4.12\times10^{9}$ & 181 \\
4.5 & 68  & 38.0 & $9.68\times10^{14}$ & 40 & 40 & $8.02\times10^{8}$ & 25 \\
5.0 & 170 & 17.3 & $1.10\times10^{15}$ & 50 & 40 & $8.27\times10^{8}$ & 25 \\
5.0 & 125 & 15.4 & $7.22\times10^{14}$ & 50 & 40 & $5.42\times10^{8}$ & 26 \\
\hline
\end{tabular}
\caption{
Proton energy ($E_p$), proton current ($I_p$), irradiation time, total protons on target ($N_p$), Li-target thickness, distance from target to DUT, PINO-calculated neutron number $N_n$ (zero incident-energy spread) through the $1.5~\mathrm{cm}^{2}$ DUT area, and observed SEUs ($N_{\text{SEU}}$). The SEU monitor was operated continuously during the experimental campaign; no SEUs were observed in the absence of beam.
}
\label{tab:CN_data}
\end{table}

Neutron numbers were obtained using PINO in the zero-spread limit. As discussed in Section~\ref{sec:modelling}, the corresponding PINO calculations reproduce the forward yields of EPEN, FLUKA, and MCNPX within $\pm5\%$ over the CN energy range.

Within the quoted uncertainties, the measurements at the three proton energies give mutually consistent device-level cross sections. The limited statistics do not permit further interpretation.

The PINO-derived neutron numbers carry systematic uncertainties of 5\% from the proton-current measurement, 4\% from DUT repositioning (1~mm), approximately 5\% from target-thickness variations, and an estimated 5\% uncertainty in the PINO yield calculation. Combined in quadrature, these contributions correspond to a $\sim$9--10\% systematic uncertainty on the neutron fluence. Statistical uncertainties from SEU counting are also included. Possible contributions from room-return neutrons were not explicitly modelled; given the short source--DUT distance (40--50~mm), the forward component is expected to dominate.

These cross sections are device-level quantities; a normalisation per bit is not provided because the internal memory organisation and error-detection chain are proprietary. The results should therefore be regarded strictly as feasibility-level indicators of device response, not as calibrated cross sections for modelling or qualification.

The test also identified several straightforward improvements for a dedicated campaign: shielding of the control circuitry, automated SEU logging to avoid double counting, fixed DUT positioning during energy scans, and accurate charge integration using an upstream Faraday cup. A Faraday cup could not be used during this parasitic run because the ongoing ToF campaign required maintaining the full beam-line geometry. This is relevant because, in the present configuration, the lithium target does not suppress secondary electrons, whose yield can reach \(\delta = 0.15\)~\cite{secondary_electrons}; this contribution was not included in the error bars of Table~\ref{tab:CN_data}. In addition, full spectral reconstruction of the CN neutron field will be required for quantitative SEE studies.

Together with the ToF verification, these SEU observations confirm that measurable upset activity can be detected under CN operating conditions. The feasibility run also identified practical upgrades—higher proton currents, improved shielding, accurate charge integration, and stable DUT positioning—required for a statistically meaningful SEU campaign.

%================= Conclusions ============
\section{Conclusions}
\label{sec:conclusions}

The present measurements establish the operational feasibility of the CN~0$^\circ$ quasi-monoenergetic neutron beam line, but do not yet provide the spectral and absolute-flux information required for a fully calibrated reference field. Within this scope, the results define the current performance of the facility and clarify the steps required for quantitative SEU studies.

Thin lithium targets provide a well-defined quasi-monoenergetic neutron component in the 0.7--4~MeV range, and the secondary pulsing system enables sub-nanosecond time-of-flight measurements. Together, these features allow direct verification of the timing structure and forward-angle kinematics of the neutron field at the detector position.

Completion of the planned shielding upgrades will allow operation at higher continuous proton currents, increasing the available neutron fluence. Routine ToF-based checks will be required to monitor spectral stability following changes in target thickness, beam tuning, or local shielding.

The feasibility-level SEU measurements confirm that measurable upset activity can be observed under CN operating conditions. While these data are not sufficient for extracting detailed device-response parameters, they demonstrate that the available neutron flux and timing structure are adequate for dedicated low-MeV SEU campaigns once higher continuous currents are available.

In summary, the CN~0$^\circ$ beam line provides a verified forward quasi-monoenergetic neutron capability in the 0.7--4~MeV range. With incremental upgrades and continued calibration work, it can support controlled low-MeV neutron studies under controlled experimental conditions.

\acknowledgments
The author would like to thank Luca Maran and Gianluca Finocchiaro for their support during CN operation and for helpful discussions on beam conditions and experimental constraints.

%================== REFERENCES ==================

\end{document}